# Massively Parallel Coincidence Counting of High-Dimensional Entangled States

Matthew Reichert[*], Hugo Defienne, and Jason W. Fleischer[†]
Department of Electrical Engineering, Princeton University, Princeton, NJ 08544, USA
[*]mr22@princeton.edu, [†]jasonf@princeton.edu

**Quantum entangled states of light are essential for quantum technologies and fundamental tests of physics. While quantum information science has relied on systems with entanglement in 2D degrees of freedom[1-4], e.g. quantum bits with polarization states, the field is moving towards ever-higher dimensions of entanglement. Increasing the dimensionality enhances the channel capacity and security of quantum communication protocols[5-10], gives rise to exponential speed-up of quantum computation[11,12], and is necessary for quantum imaging[10,13-18]. Yet, characterization of even bipartite quantum states of high-dimensional entanglement remains a prohibitively time-consuming challenge[19,20], as the dimensionality of the joint Hilbert space scales quadratically with the number of modes. Here, we develop and experimentally demonstrate a new, more complete theory of detection in CCD cameras for rapid measurement of the full joint probability distribution of high-dimensional quantum states. The theory spans the intensity range from low photon count to saturation of the detector, while the massive parallelization inherent in the pixel array makes measurements scale favorably with dimensionality. The results accurately account for partial detection and electronic noise, resolve the paradox of ignoring two-photon detection in a single pixel despite collinear spatial entanglement, and reveal the full Hilbert space for exploration. For example, use of a megapixel array allows measurement of a joint Hilbert space of $10^{12}$ dimensions, with a speed-up of nearly four orders of magnitude over traditional methods. We demonstrate the method with pairs, but it generalizes readily to arbitrary numbers of entangled photons. The technique uses standard geometry with existing technology, thus removing barriers of entry to quantum imaging experiments, and open previously inaccessible regimes of high-dimensional quantum optics.**

Broad beams of quantum light are a natural pathway to large Hilbert spaces, as they have high-dimensional entanglement in transverse spatial modes[13]. Spatial correlation of biphotons has led to sub-shot-noise quantum imaging[14,15], enhanced resolution[10], quantum ghost imaging[16], and proposals for quantum lithography[17]. Despite this work, high-dimensional quantum optics remains underdeveloped, largely due to difficulty in measuring the full joint probability distribution. Traditionally, experiments measure coincidences between two single-photon counting modules (SPCMs) that are each scanned over their own subspace to build up a measurement point-by-point. Such a procedure is photon-inefficient, making high-dimensional measurements tedious and prohibitively time consuming. Full quantum-state measurements are impractical even for a relatively small number of dimensions[19,20].

In this work, we present a rapid and efficient method of measuring a high-dimensional biphoton joint probability distribution via massively parallel coincidence counting. We use a single-photon-sensitive electron-multiplying (EM) CCD camera as a dense array of photon detectors to measure all dimensions of the joint Hilbert space simultaneously. For example, a typical megapixel camera can record a one trillion-dimensional joint Hilbert space nearly 10,000× faster than traditional raster-scanning methods. This speed-up enables direct access to high-dimensional spaces that are impractical to measure through standard means.

Recent efforts with single-photon-sensitive cameras have characterized spatial entanglement[18,21-25], but results relied on projection onto two dimensions and have been limited to demonstrations of EPR-type entanglement of homogeneous distributions. Furthermore, to mitigate complications of accidental counts, coincidence measurements were performed in the low-count-rate regime. In this case, the coincidence count rate is assumed proportional to the biphoton joint probability distribution (atop a noise baseline). Here, we show that this assumption is unnecessary and give a general expression for the biphoton joint probability distribution. The exact expression follows from measurements of single- and coincidence-count probabilities and is valid for arbitrary count rates up to detector saturation, enabling more accurate measurements, faster acquisition speeds, and optimization of the signal-to-noise ratio.

To demonstrate our method, we characterize the properties of photon pairs entangled in transverse spatial degrees of freedom. Like classical light-field methods[26], measurement of the full 4D distribution shows details and features that would be lost with traditional projection methods. While we consider transverse spatial degrees of freedom, we emphasize that our technique may be readily extended to other degrees of freedom, such as spectral modes or orbital angular momentum, by suitable mapping onto the pixels of the camera.

The transverse spatial dependence of a pure entangled photon state is described by the biphoton wave function $\psi(\boldsymbol{\rho}_i, \boldsymbol{\rho}_j)$, where $\boldsymbol{\rho}_i = x_i \hat{\mathbf{x}}_i + y_i \hat{\mathbf{y}}_i$, and likewise for $\boldsymbol{\rho}_j$. The joint probability of observing one photon at $\boldsymbol{\rho}_i$ and its pair at $\boldsymbol{\rho}_j$ is $\Gamma(\boldsymbol{\rho}_i, \boldsymbol{\rho}_j) = |\psi(\boldsymbol{\rho}_i, \boldsymbol{\rho}_j)|^2$, which in a discretized basis is $\Gamma_{ij}$. Since each photon may be found in a 2D space ($x_i$, $y_i$), the joint probability distribution is a 4D distribution. A schematic of the measurement and processing procedure is shown in Figure 1. Spatially entangled photon pairs are generated via spontaneous parametric down-conversion (SPDC) in a β-barium borate (BBO) crystal, cut for type-I phase matching. The crystal is pumped by a 120 mW, 400 nm cw laser diode that is spatially filtered and collimated (not shown). Spectral filters block the pump beam and select near-degenerate photon pairs at 800 nm (40 nm FWHM). These are placed immediately after the BBO crystal to prevent induced fluorescence in the subsequent optics. A lens images the far field of the crystal onto an EMCCD camera (Andor iXon Ultra).

Measurement of the biphoton joint probability distribution $\Gamma_{ij}$ is possible with an EMCCD camera due to its high quantum efficiency and low noise floor. The camera is operated in photon-counting regime, where each pixel is set to one if its gray level output is above a threshold and zero otherwise[27] (see Methods). The data consist of a set of $N$ frames $C_{i,n} = \{0,1\}$, where subscript $i$ is the pixel index (spatial mode) and $n$ is the frame number. Each frame consists of many counts from both photon events and electronic noise (mainly due to clock-induced charge[27]). A measure of the count probability can be obtained by an average over all frames

$$\langle C_i \rangle = \sum_m P_m \left( \mu_{i|m} + p_{el} \mu_{\bar{i}|m} \right), \tag{1}$$

where $P_m$ is the distribution of the number $m$ of photon pairs and $p_{el}$ is the electronic count probability (e.g., dark counts). The factors $\mu_{i|m}$ and $\mu_{\bar{i}|m}$ represent the conditional probabilities of detecting *at least* one photon and zero photons, respectively, given $m$ pairs arriving within the detector time window (see Table 1)[28].

The duration of both exposure and read-out of each frame of the EMCCD is much longer than the biphoton correlation time. Therefore, photons from each pair arrive at the camera within a single frame. Coincidences are measured by the average of the tensor product of each frame with itself

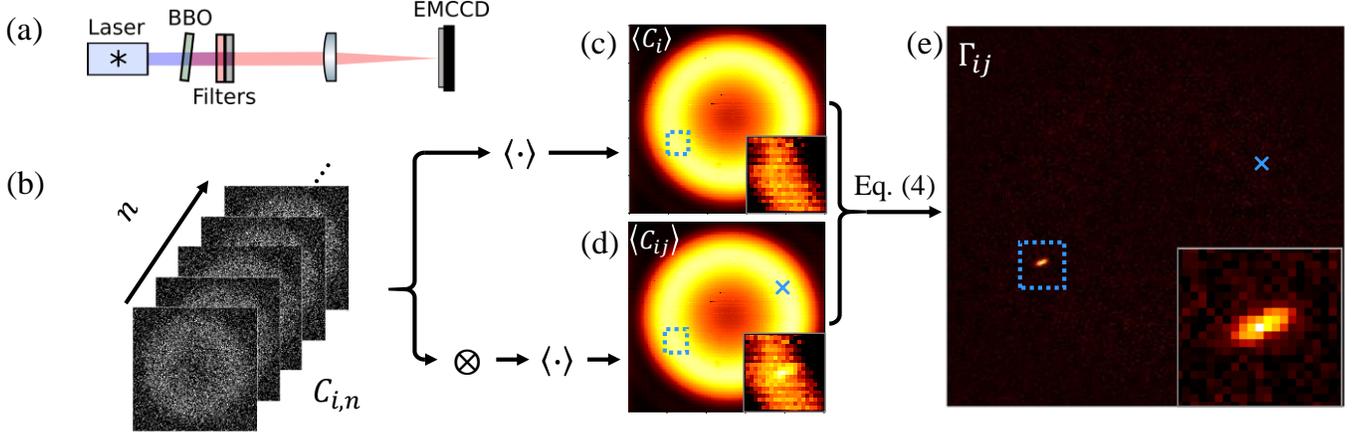

**Figure 1 | Measuring the biphoton joint probability distribution with an EMCCD camera.** (a) Experimental setup for measuring far-field type-I SPDC. (b-e) Flow chart of data processing. (b) The camera acquires many (~ $10^5 - 10^7$) thresholded frames from which both the (c) average of all frames $\langle C_i \rangle$ (indicated by $\langle \cdot \rangle$) and (d) average of the tensor product of each frame with itself ($\otimes$, Eq. (2)) are calculated. The latter gives the mean coincidence probability $\langle C_{ij} \rangle$. (d) A 2D slice of the 4D $\langle C_{ij} \rangle$ for $j = [x_j = 70, y_j = 33]$ (as indicated by the blue x) appears very similar to $\langle C_i \rangle$, since most of the coincidences are accidentals between photon from different pairs. Genuine coincidences from anti-correlated entangled photons appear within the boxed region (see insets). (e) The conditional probability distribution $\Gamma_{i|j}$, calculated from $\Gamma_{ij}$ via Eq. (4), shows that paired photons are localized about $i = [-70, -32]$.

$$\langle C_{ij} \rangle = \frac{1}{N} \sum_{n=1}^{N} C_{i,n} C_{j,n}. \quad (2)$$

This is a measure of the coincidence count probability between all pixels $i$ and $j$. In addition to genuine coincidence counts from entangled photon pairs, there are also accidental counts between uncorrelated photons and noise. These can be accounted for in general by the expression

$$\langle C_{ij} \rangle = \sum_m P_m \left( \mu_{ij|m} + p_{el}(\mu_{i\bar{j}|m} + \mu_{\bar{i}j|m}) + p_{el}^2 \mu_{\bar{i}\bar{j}|m} \right), \quad (3)$$

where each of the terms $\mu_{pq|m}$ are related to the joint probability distribution and its marginal (see Table 1)[28]. The three terms in Eq. (3) are coincidences between 1) *at least* two photons, 2) *at least* one photon and one electronic noise event, and 3) two noise events. For a Poissonian number distribution of generated photon pairs, the summation in Eq. (3) simplifies, giving an analytic expression for $\langle C_{ij} \rangle$ in terms of $\langle C_i \rangle$, $\langle C_j \rangle$, and $\Gamma_{ij}$. This may be solved with Eq. (1) to yield

$$\Gamma_{ij} = \alpha \ln \left( 1 + \frac{\langle C_{ij} \rangle - \langle C_i \rangle \langle C_j \rangle}{(1 - \langle C_i \rangle)(1 - \langle C_j \rangle)} \right) \quad (4)$$

where $\alpha$ is a constant that depends on the quantum efficiency of the system (see Methods).

Note that Eq. (4) includes the case when several photons arrive at the same pixel. This case has been excluded explicitly by other treatments, even though collinear geometry and high spatial entanglement make this case the most likely one. The paradox is often circumvented by considering the low-photon-count limit, in which the joint probability distribution $\Gamma_{ij}$ becomes proportional to the measured coincidence count rate $\langle C_{ij} \rangle$. However, this assumption is not necessary here. Indeed, Eq. (4) remains valid up to detector saturation. The formalism thus covers the entire range of photon intensities and types of detection events, and generalizes straight forwardly to joint distributions of higher numbers of entangled photons.

Figure 1d shows the coincidence count distribution for a particular pixel $j = [x_j = 70, y_j = 33]$, i.e., a 2D slice for all $i = \{x_i, y_i\}$ through the 4D joint distribution $\langle C_{ij} \rangle$. It includes genuine coincidences as well as a large background from accidental counts. Due to the large number of pairs in each frame (~$10^4$), most accidentals are between photons from different pairs; indeed, Figure 1d appears very similar to the singles count distribution $\langle C_i \rangle$ in Figure 1c. Genuine coincidences between photons from the same pair, shown in the inset, rise above the background from accidentals. The corresponding 2D slice through the 4D $\Gamma_{ij}$, calculated via Eq. (4), is displayed in Figure 1e. When one photon is found at $j = [70, 33]$, its entangled partner is localized near $i = [-70, -32]$, indicating a high degree of anti-correlation. Such conditional distributions $\Gamma_{i|j}$ are measured simultaneously for all $j$, thus constituting a full measurement of the 4D biphoton joint probability distribution.

**Table I** Probabilities of single point detection $\mu_{p|m}$ and coincidence $\mu_{pq|m}$ conditioned on the number of generated photon pairs $m$. $\Gamma_{pq}$ is the joint probability distribution, $\Gamma_p$ is the marginal, and $\eta$ is the detection quantum efficiency. Barred subscript indicates no detection.

| Term | Expression |
|---|---|
| $\mu_{i|m}$ | $1 - \mu_{\bar{i}|m}$ |
| $\mu_{\bar{i}|m}$ | $(1 - 2\eta\Gamma_i + \eta^2 \Gamma_{ii})^m$ |
| $\mu_{ij|m}$ | $1 - \mu_{\bar{i}|m} - \mu_{\bar{j}|m} + \mu_{\bar{i}\bar{j}|m}$ |
| $\mu_{i\bar{j}|m}$ | $\mu_{\bar{j}|m} - \mu_{\bar{i}\bar{j}|m}$ |
| $\mu_{\bar{i}\bar{j}|m}$ | $\left(1 - 2\eta(\Gamma_i + \Gamma_j) + \eta^2(\Gamma_{ii} + \Gamma_{jj} + \Gamma_{ij})\right)^m$ |

Complete measurements of high-dimensional joint Hilbert spaces contain detailed, localized (i.e., non-averaged) information about correlations of entangled photon pairs. Figures 2a-c show $\Gamma_{i|j}$ for conditional photons detected at different radial distances $j = [x_j, y_j]$ from the center of the beam. As $x_j$ is increased, $x_i$ decreases to maintain a fixed sum, i.e., $x_i + x_j \approx 0$. However, there is also a variation in the shape of the conditional distributions themselves: the width along the radial direction increases with $x_i$. This arises from the radial dependence of the uncertainty in the wave vector **k**, $\Delta k_\rho \approx k_\rho |\Delta \mathbf{k}|/|\mathbf{k}|$. Observation of such features with traditional raster-scanning techniques requires multiple separate measurements. With an EMCCD camera, they are all captured simultaneously in a single image.

In previous studies, measurements of the intercorrelation function have been demonstrated via image correlation techniques[21,22], without measuring the full 4D $\Gamma_{ij}$. However, such measurements provide only the globally averaged correlation and thus neglect any potential internal variation in the joint probability distribution. To show this, we project $\Gamma_{ij}$ onto the coordinate sums $[(x_i + x_j)/\sqrt{2}, (y_i + y_j)/\sqrt{2}]$ (Figure 2d), which is the spatially averaged intercorrelation function in momentum space. The peak near the center indicates that entangled photon pairs are always found near equal and opposite sides of the center, within anti-correlation widths $\sigma_{y,+} = 18.6 \pm 0.3$ μm and $\sigma_{x,+} = 20.9 \pm 0.3$ μm. Our more-resolved methods show that, even in this simple case, the corresponding widths of the $\Gamma_{i|j}$ in Figures 2a-c vary significantly, with $\sigma_x = 16.1 \pm 1.4$ μm, $23.0 \pm 1.5$ μm, and $34.9 \pm 2.5$ μm, respectively.

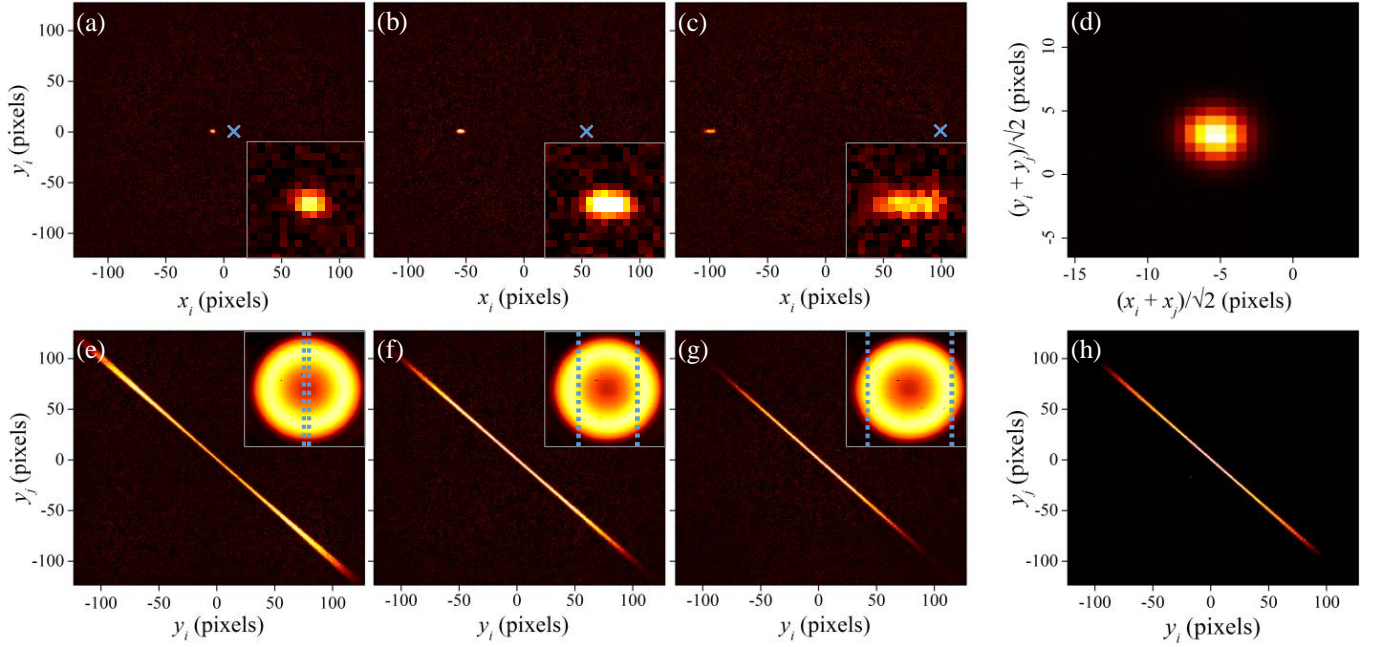

**Figure 2 | Information contained in the full 4D measurement of biphoton joint probability distribution.** (a-c) Variation of $\Gamma_{i|j}$ at different distances from the center—indicated by blue x—showing anti-correlation and increasing correlation width. (d) Projection of $\Gamma_{ij}$ onto sum and difference coordinates averages the variations in (a-c). (e-g) 2D slices of $\Gamma_{ij}$ for fixed $[x_i, x_j]$ (as indicated by blue dashed lines in inset of $\langle C_i \rangle$) showing variation in anti-correlation with horizontal separation. (h) Projection of $\Gamma_{ij}$ onto $[y_i, y_j]$ (integration over $x_i$ and $x_j$) averages the structures in (e-g), giving only a mean profile.

Other slices of $\Gamma_{ij}$, along different coordinates, contain different information about the entangled photon pairs. For example, rather than fixing $[x_j, y_j]$ to see $\Gamma_{i|j}$, we may examine correlations in vertical position within specific columns of the image by fixing $[x_i, x_j]$. Examples in Figures 2e-g show strong anti-correlation of variable width, each taken at different reference columns (indicated in the insets). We observe strong vertical anti-correlation that changes depending on the horizontal separation of the selected columns. The radial variation of $\Gamma_{i|j}$ in Figures 2a-c gives a $|y|$-dependence of vertical anti-correlation, which diminishes for larger $|x|$. Projecting $\Gamma_{ij}$ (Figure 2h) averages this variation, resulting in lost information.

The massively parallel capability of EMCCD cameras allows for much faster measurement of joint probability distributions than traditional scanning techniques. Raster-scanning pairs of SPCMs, each in a $d$-dimensional plane, requires $d^2$ measurements to build a complete measurement. In contrast, an EMCCD measures the entire plane at once, with pixels at each point in the array. While SPCMs have a high effective frame rate (10s of MHz), the acquisition time of an EMCCD camera is practically limited by the readout process. This affects the camera frame rate, which, for a square frame, scales as $\sqrt{d}$. For the camera used here, a definitive speed advantage is found for $d^2 > (24 \times 24)^2 \approx 330{,}000$ (see Supplement). Use of the full frame allows measurement of up to $(1024 \times 1024)^2 \approx$ one trillion dimensional joint Hilbert space. At 26 frames per second, measurements of $\Gamma_{ij}$ could achieve a signal-to-noise ratio of 10 in approximately 11 hours. The same measurement performed with raster-scanning SPCMs is estimated to take 9 years, giving a camera improvement of ~ 7000×. Data shown in Figures 1 and 2 were taken from a subset of $251 \times 251$ pixels, corresponding to a four-billion-dimensional joint Hilbert space, and were acquired in a matter of hours. The EMCCD camera also outperforms compressive sensing methods[29] for large joint Hilbert spaces and does not require sparsity or numerical retrieval.

Camera-based methods hold clear advantages for quantum imaging applications, which offer improved performance over systems using classical coherent light. Imaging with perfectly correlated photon pairs—with biphoton wave function $\psi(\boldsymbol{\rho}_i, \boldsymbol{\rho}_j) = \delta(\boldsymbol{\rho}_i - \boldsymbol{\rho}_j)$—gives a probability distribution of both photons at the same position in the image plane

$$\Gamma(\boldsymbol{\rho}, \boldsymbol{\rho}) \propto \left| \int t^2(\boldsymbol{\rho}') h^2(\boldsymbol{\rho} - \boldsymbol{\rho}') d\boldsymbol{\rho}' \right|^2 \quad (5)$$

where $t(\boldsymbol{\rho})$ is the object transmittance and $h(\boldsymbol{\rho})$ is the point spread function. The fact that the square of $h(\boldsymbol{\rho})$ appears in Eq. (5) means that biphoton imaging has higher resolution than conventional laser imaging[17,30] (see Methods).

As an example, we image a standard USAF resolution chart with entangled photon-pair illumination (Figure 3). Unlike the previous examples using anti-correlated biphotons, here we use spatially correlated biphotons—where one photon is localized near its partner ($i \approx j$)—by projecting the output facet of the nonlinear crystal onto the object, which is then imaged onto the camera. An adjustable iris is placed in a Fourier plane to control the numerical aperture. To ensure the validity of Eq. (5), we measure the incident $\Gamma_{ij}$ without the object; the results confirm strong spatial correlation, visible in both the conditional distributions (Figures 3b,c) and the projection onto the difference coordinates (Figures 3d). By fitting to a Gaussian distribution, we find the correlation width $\sigma_- = 8.5 \pm 0.5$ μm. Measurements are then repeated with the object; a 3D projection of $\Gamma_{ij}$ (Figure 3e) displays the image of the resolution chart, its appropriate basis (diagonal plane), and the final spatial correlation distribution of the biphotons (width of the diagonal plane). Coincidence images taken with entangled photon pairs (Figure 3f) show clear improvement in resolution over those with a 808 nm laser diode (Figure 3g), with less noise and higher visibility. For example, the bars within the red boxed region (group 4, element 6) are clearly resolved with quantum light (visibility of $0.33 \pm 0.03$), but not with classical coherent light (visibility < 0.04).

By using readily available technology and standard imaging geometries, our method removes barriers of entry to experiments in quantum optics. Time-resolved measurements of coincidence counts are replaced by time-averaged camera measurements of photon correlations, while lower-order counts and conditional probabilities are bootstrapped to provide complete characterization of joint distribution functions. Further, the massive parallelization inherent in megapixel cameras enables measurement of states with orders-of-magnitude greater dimensionality than previously possible, with similar increases in acquisition speed. With suitable mapping for other degrees of freedom, e.g. dispersive elements for spectral modes or diffractive elements for orbital angular momentum, other types of quantum states can be characterized as well. The results thus extend conventional imaging to the quantum domain, providing a pathway for quantum phase retrieval and coherence/entanglement control, and enable new means of quantum information processing with high-dimensional entangled states.

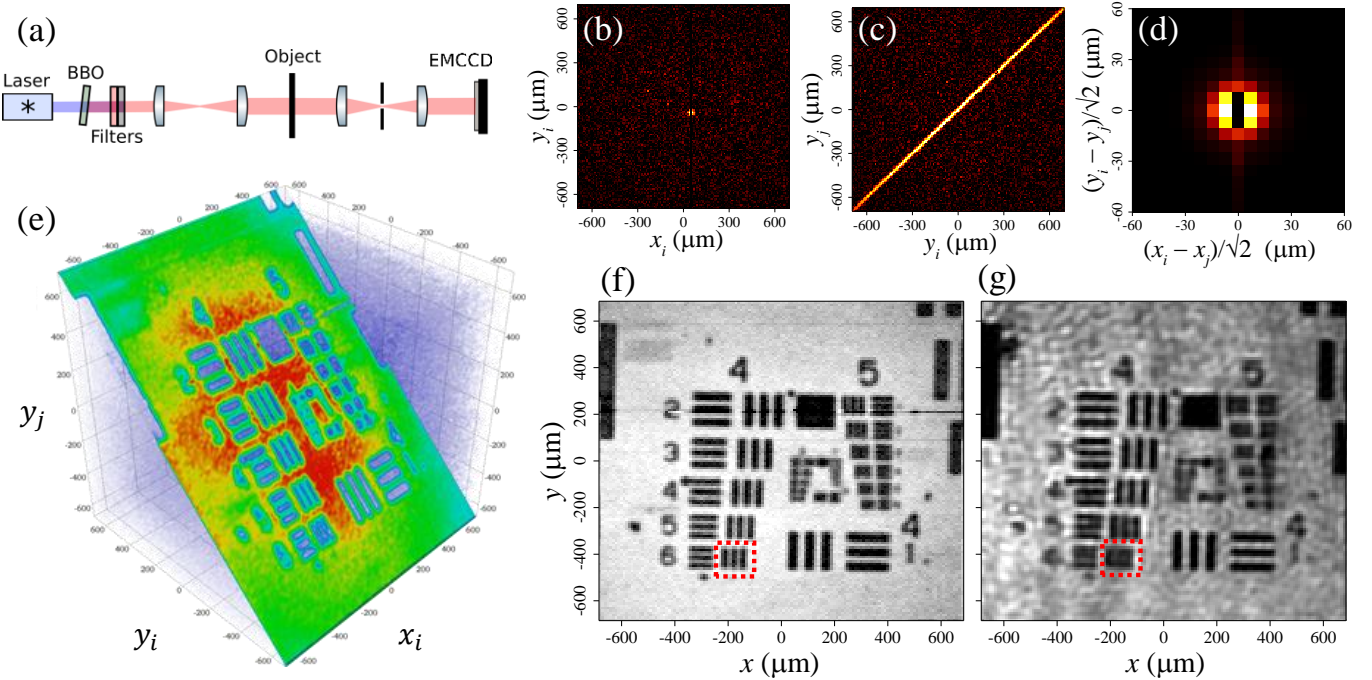

**Figure 3 | Biphoton imaging of a USAF resolution chart with an EMCCD camera** (a) Experimental setup for biphoton imaging. (b-d) Measurements of incident $\Gamma_{ij}$ (without the object), showing (b) $\Gamma_{i|j}$ for $j = [x_i = 50$ μm, $y_i = –40$ μm], (c) 2D slice of $\Gamma_{ij}$ for fixed $[x_i, x_j]$, and (d) projection onto the difference coordinates. Each shows a high degree of spatial correlation. Black region $x_j = x_i$ in (b, d) results from zeroing to eliminate the artifact from charge transfer inefficiency (see Methods and Supplement). (e) 3D projection of $\Gamma_{ij}$ onto $(x_i, y_i, y_j)$, simultaneously shows the image of the resolution chart in the appropriate basis (diagonal plane) and spatial correlation between $y_i$ and $y_j$. (f, g) Comparison of 2D image of resolution chart taken with (f) entangled photon pairs and (g) classical coherent light (808 nm laser diode) at NA ~ 0.016.

## METHODS

The EMCCD (iXon Ultra 897, Andor) is a highly sensitive camera where an avalanche gain of up to 1000 amplifies the signal in each pixel before readout. The camera has a pixel size of 16×16 μm$^2$ with a quantum efficiency of ~70 % at 800 nm. To minimize the dark-count rate compared to other noise sources in the camera, it is operated at a temperature of –85 °C. The camera is first characterized by measuring the histogram of the gray scale output of each pixel from many (~10$^6$) frames taken with the shutter closed. The histogram is primarily Gaussian, due to read noise, with an additional exponential tail towards high gray levels due primarily to clock-induced charge (CIC) noise[27]. We fit the histogram with a Gaussian distribution to find the center (~170) and standard deviation $\sigma$ (4 to 20, depending on the readout rate). We have found that a threshold set to $2\sigma$ above the mean maximizes the signal-to-noise ratio. A pixel-dependent threshold is used to account for a minor inhomogeneity across the frame. There is a small cross talk effect between pixels in a single column due to sub-optimal charge transfer efficiency upon readout (see Supplement). For this reason, within each 2D frame of $\Gamma_{i|j}$, we set to zero the 10 pixels above and below $i = j$.

Operating at higher readout rate increases readout and CIC noise, but we have found that the increased acquisition rate more than compensates, yielding a higher signal-to-noise ratio (SNR) for the same total acquisition time. The camera is therefore operated at the fastest available settings: a horizontal readout rate of 17 MHz and a vertical shift time of 0.3 μs, with a vertical clock voltage of +4 V. The pump laser power and camera exposure time are set to give an optimum peak count probability $\langle C \rangle$ of ~0.2[27]. We acquire a number of frames sufficient to achieve the desired SNR. Typically, a series of ~10$^5$-10$^7$ images are acquired at a ~1-5 ms exposure time. Many sets of thresholded frames are saved to disk, where each set contains 10$^4$ frames as a logical array $C_{i,n}$. Each column of the array represents a single frame, and each row represents a pixel. Eq. (2) is used to calculate $\langle C_{ij} \rangle$ by matrix multiplication of each set of frames, which are then averaged. To minimize non-ergodic effects, the term $\langle C_i \rangle \langle C_j \rangle$ in Eq. (4) is calculated via matrix multiplication of successive frames (see Supplement). Elsewhere, $\langle C_i \rangle$ is the average of all frames.

In general, the biphoton wave function in an image plane is given by

$$\psi_{img}(\boldsymbol{\rho}_i, \boldsymbol{\rho}_j) = \iint h(\boldsymbol{\rho}_i - \boldsymbol{\rho}'_i) h(\boldsymbol{\rho}_j - \boldsymbol{\rho}'_j) \cdot t(\boldsymbol{\rho}'_i) t(\boldsymbol{\rho}'_j) \psi_s(\boldsymbol{\rho}'_i, \boldsymbol{\rho}'_j) d\boldsymbol{\rho}'_i d\boldsymbol{\rho}'_j \quad (6)$$

where $\psi_s(\boldsymbol{\rho}_i, \boldsymbol{\rho}_j)$ is the wave function incident on the object. With ideally correlated photon pairs, i.e., $\psi_s(\boldsymbol{\rho}_i, \boldsymbol{\rho}_j) = \delta(\boldsymbol{\rho}_i - \boldsymbol{\rho}_j)$, the square amplitude of Eq. (6) simplifies to Eq. (5). The high-resolution biphoton image therefore lies within $\Gamma_{ii}$, where both entangled photons hit the same pixel. However, as EMCCDs are not photon-number-resolving, it cannot distinguish between one or both photons hitting the same pixel. Instead, we approximate $\Gamma_{ii}$ by the case where the two entangled photons arrive in adjacent pixels, i.e., $\Gamma_{i,i+1}$, as we do in Figure 3f. This assumption is valid when the biphoton correlation width and image features are both larger than the pixel size.

For ideal imaging ($h(\boldsymbol{\rho}) \approx \delta(\boldsymbol{\rho})$), intensity images are directly proportional to $|t(\boldsymbol{\rho})|^2$, where $t(\boldsymbol{\rho})$ is the complex (field) function for transmission. For entangled-photon images, $\Gamma(\boldsymbol{\rho}, \boldsymbol{\rho}) \propto |t(\boldsymbol{\rho})|^4$ (see Eq. (5)). Therefore, we show in Figures 3f,g the intensity images of coherent-state radiation and the square root of the biphoton images.

**Acknowledgements** The authors would like to thank Nova Photonics, Inc. for providing equipment used in the experiment. This work was supported by the Air Force Office of Scientific Research grant FA9550-12-1-0054.

**Author contributions** M.R. and J.W.F. conceived of the experiment; M.R. and H.D. developed the theory and experimental design, and M.R. performed the experiment; all authors analyzed the data and co-wrote the paper.

# Massively Parallel Coincidence Counting of High-Dimensional Entangled States: Supplement

Matthew Reichert[*], Hugo Defienne, and Jason W. Fleischer[†]
Department of Electrical Engineering, Princeton University, Princeton, NJ 08544, USA
[*]mr22@princeton.edu, [†]jasonf@princeton.edu

## 1. Biphoton Joint Probability Distribution

Experimentally, there are two possible ways of measuring entangled photon pairs with detector arrays: (a) photons from pairs are deterministically separated to different detector arrays (or different regions of a single array), and (b) photons are all sent to a single detector array. The principle difference is that both photons from a single pair may hit the same pixel in (b) but not in (a). Equations for the two cases are presented below in tandem, and labeled a and b accordingly, along with equations shared by both cases. Note that in the following we omit the quantum efficiency $\eta$ for brevity. To incorporate it, make the substations $\Gamma_i \to \eta \Gamma_i$ and $\Gamma_{ij} \to \eta^2 \Gamma_{ij}$.

The singles count probability at pixel $i$ is given by

$$\langle C_i \rangle = \sum_m P_m \left( \mu_{i|m} + p_{el} \mu_{\bar{i}|m} \right), \tag{S1}$$

where $P_m$ is the probability distribution for the number of generated pairs and $p_{el}$ is the electronic count probability of the detector (dark counts, CIC, etc.). $\mu_{i|m}$ is the conditional probability, given $m$ photon pairs, that *at least* one photon is detected in pixel $i$. $p_{el}\mu_{\bar{i}|m}$ is the probability of electronic noise counts (counts not due to photons), which requires the absence of detected photons. The factor $\mu_{\bar{i}|m}$ is the conditional probability, given $m$ photon pairs, that no photons are detected in pixel $i$ (indicated by the barred $i$), which is related to the marginal distribution by

$$\mu_{\bar{i}|m} = (1 - \Gamma_i)^m, \tag{S2a}$$

$$\mu_{\bar{i}|m} = \left(1 - (2\Gamma_i - \Gamma_{ii})\right)^m. \tag{S2b}$$

In case (a), only one photon from the pair is sent to the detector array of mode $i$, while in case (b) both photons from the pair go to the same array. Multiplying $\Gamma_i$ by 2 accounts for this, but $\Gamma_i$ also includes the case where the other photon is also in mode $i$, i.e., $\Gamma_{ii}$. Doubling $\Gamma_i$ double counts this occurrence, and therefore we must subtract off the extra factor of $\Gamma_{ii}$.

Because $\mu_{i|m}$ and $\mu_{\bar{i}|m}$ sum to unity, they are related by

$$\mu_{i|m} = 1 - \mu_{\bar{i}|m}. \tag{S3}$$

For a Poissonian number distribution of pairs, $P_m = \langle m \rangle^m e^{-\langle m \rangle}/m!$, where $\langle m \rangle$ is the mean number of photon pairs emitted within exposure time $\tau_e$ [1,2], Eq. (S1) simplifies to

$$\langle C_i \rangle = 1 - (1 - p_{el}) e^{-\langle m \rangle \Gamma_i}, \tag{S4a}$$

$$\langle C_i \rangle = 1 - (1 - p_{el}) e^{-\langle m \rangle (2\Gamma_i - \Gamma_{ii})}. \tag{S4b}$$

The coincidence count probability between pixels $i$ and $j$ is

$$\langle C_{ij} \rangle = \sum_m P_m \left( \mu_{ij|m} + p_{el} \left( \mu_{i\bar{j}|m} + \mu_{\bar{i}j|m} \right) + p_{el}^2 \mu_{\bar{i}\bar{j}|m} \right), \tag{S5}$$

where the first term represents the probability of coincidence between two photons, the second between one photon and one electronic noise event, and the third between two noise events. The sum of the $\mu$'s is unity: $\mu_{ij|m} + \mu_{i\bar{j}|m} + \mu_{\bar{i}j|m} + \mu_{\bar{i}\bar{j}|m} = 1$. Coincidences between two electronic noise events depend on photon detections in either pixel $i$ or $j$, which is given by [3]

$$\mu_{\bar{i}\bar{j}|m} = \left(1 - \Gamma_i - \Gamma_j + \Gamma_{ij}\right)^m, \quad (S6a)$$

$$\mu_{\bar{i}\bar{j}|m} = \left(1 - (2\Gamma_i - \Gamma_{ii}) - (2\Gamma_j - \Gamma_{jj}) + 2\Gamma_{ij}\right)^m. \quad (S6b)$$

Coincidence counts between photons and electronic noise requires *at least* one photon detection in one pixel and zero in the other. This is given by the probability that no photons are detected in one pixel, i.e., $\mu_{\bar{j}|m}$, minus the probability that no photons are detected in either pixel, $\mu_{\bar{i}\bar{j}|m}$, that is

$$\mu_{i\bar{j}|m} = \mu_{\bar{j}|m} - \mu_{\bar{i}\bar{j}|m}, \quad (S7)$$

and vice-versa for $\mu_{\bar{i}j|m}$. The probability that *at least* one photon is detected in each pixel $i$ and $j$ is then

$$\mu_{ij|m} = 1 - \mu_{\bar{i}|m} - \mu_{\bar{j}|m} + \mu_{\bar{i}\bar{j}|m}. \quad (S8)$$

For a Poissonian number distribution of generated pairs, Eq. (S5) becomes

$$\langle C_{ij} \rangle = 1 - (1 - p_{el})\left(e^{-\langle m \rangle \Gamma_i} + e^{-\langle m \rangle \Gamma_j}\right) + (1 - p_{el})^2 e^{-\langle m \rangle (\Gamma_i + \Gamma_j - \Gamma_{ij})}, \quad (S9a)$$

$$\langle C_{ij} \rangle = 1 - (1 - p_{el})\left(e^{-\langle m \rangle (2\Gamma_i - \Gamma_{ii})} + e^{-\langle m \rangle (2\Gamma_j - \Gamma_{jj})}\right) + (1 - p_{el})^2 e^{-\langle m \rangle \left((2\Gamma_i - \Gamma_{ii}) + (2\Gamma_j - \Gamma_{jj}) - 2\Gamma_{ij}\right)}. \quad (S9b)$$

Eqs. (S4) and (S9) can thus be used to solve for $\Gamma_{ij}$:

$$\Gamma_{ij} = \alpha \ln\left[1 + \frac{\langle C_{ij} \rangle - \langle C_i \rangle \langle C_j \rangle}{(1 - \langle C_i \rangle)(1 - \langle C_j \rangle)}\right], \quad (S10)$$

where

$$\alpha = \frac{1}{\langle m \rangle \eta^2}, \quad (S11a)$$

$$\alpha = \frac{1}{2\langle m \rangle \eta^2}. \quad (S11b)$$

Therefore, to within a constant scaling factor, only the mean coincidence- and singles-count probabilities are necessary to uniquely extract the joint probability distribution.

## 2. Residual Background

In principle, the quantity $\langle C_i \rangle \langle C_j \rangle$ may be calculated as the product of the average of all collected frames. However, we have found that doing so results in a residual background that is not due to genuine coincidence counts. For example, the black data in Fig. S1 shows the momentum anti-correlation measurement in the far-field when the term $\langle C_i \rangle \langle C_j \rangle$ in Eq. (4) is given by

$$\langle C_i \rangle \langle C_j \rangle = \left(\frac{1}{N}\sum_{n=1}^{N} C_{i,n}\right)\left(\frac{1}{N}\sum_{n'=1}^{N} C_{j,n'}\right) \quad (S12)$$

However, since these ensemble averages are calculated via temporal averages, the measurements are susceptible to non-ergodicity. In addition to potential fluctuations in pump laser power, we believe this is due to long-term fluctuation of the gain of the EMCCD camera.

The probability of getting a gray level above threshold, $P(x_g > T | n)$, depends on the input number of photoelectrons $n$, as well as on the noise and gain properties of the camera. The coincidence count distribution depends on

$$P(x_{g,i} > T, x_{g,j} > T | n_i, n_j), \tag{S13}$$

which is the probability of getting gray levels above threshold at both pixels $i$ and $j$ given $n_i$ and $n_j$ photoelectrons at the inputs. (Note that this function depends on the gain of the EMCCD camera [4,5].) Since the pixels are not correlated, the conditional probability factorizes:

$$P(x_{g,i} > T, x_{g,j} > T | n_i, n_j) = P(x_{g,i} > T | n_i) P(x_{g,j} > T | n_j). \tag{S14}$$

Long-time fluctuations of the gain cause variations in the conditional probability distribution over the course of data acquisition, which means that the temporal average does not factorize.

If we instead approximate $\langle C_i \rangle \langle C_j \rangle$ as the product of one frame with the next, i.e.,

$$\langle C_i \rangle \langle C_j \rangle \approx \frac{1}{N-1} \sum_{n=1}^{N-1} C_{i,n} C_{j,n+1}, \tag{S15}$$

then the residual background is nearly eliminated, (black data in Fig. S1c) This approximation is justified by the fact that successive frames do not contain photons from the same pair. That is, the acquisition time is sufficiently long—the inverse of the frame rate is orders of magnitude larger than the biphoton correlation time—that genuine coincidences between pairs of photons only ever occur within a single frame and never between different frames.

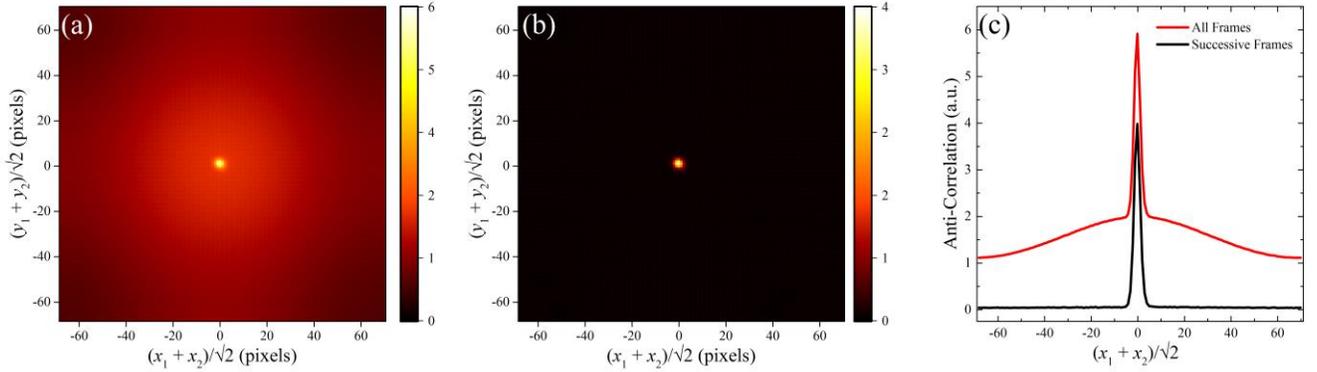

**Fig. S1 | Residual background elimination.** Measurement of momentum anti-correlation calculated via Eq. (4) from main text, where $\langle C_i \rangle \langle C_j \rangle$ is calculated from (a) product of the mean of all frames— Eq. (S12) —and (b) mean of product of successive frames—Eq. (S15). (c) Lineouts at $(y_1 + y_2)/\sqrt{2} = 0$ from (a) (red) and (b) (black), respectively.

### 3. Comparison to raster scanning

The detector array size of an Andor iXon Ultra 888 is $1024 \times 1024$ pixels, which would correspond to a Hilbert space of $(1024 \times 1024)^2 \approx 1$ trillion dimensions. According to Andor [6], the camera can operate at 26 frames per second when acquiring the entire frame. In far-field measurements of the 4D $\Gamma_{ij}$ with a correlation with $\sigma_+$ of 1.2 pixels, we have demonstrated a signal-to-noise ratio (SNR) of 10 with as few as $10^6$ frames of the EMCCD camera. At this SNR, we can measure a $2^{40}$-dimensional joint Hilbert space in 11 hours.

For comparison to raster-scanning single-photon-counters, we first assume ideal conditions (negligible dark count rate and unit quantum efficiency), and determine the number of "frames" we need to acquire to achieve the same SNR = 10 for the same $\Gamma_{ij}$ as above (where $\sigma_+ = 1.2$ pixels). That is, we solve

$$\text{SNR} = \frac{\sqrt{N}}{2\pi\sigma_+^2}, \tag{S16}$$

and find $N = 8187$. The acquisition time depends on the dead time of the photon counting modules, whose inverse gives the effective frame rate. For a typical single-photon-counting module (SPCM-AQRH Series, Excelitas Technologies [7]) the output count rate before saturation, i.e., the effective frame rate, is $R_f = 37$ MHz. It therefore takes $N/R_f = 220$ µs to acquire a sufficient number of frames to achieve the desired SNR. This, however, must be repeated over the entire trillion-dimensional Hilbert space. In practice, we may take advantage of the fact that the biphoton joint probability distribution is symmetric upon exchange, and reduce the number of measurements down to half a trillion. Therefore, measurement with raster-scanning point detectors would take at least (220 µs)($1024^4/2$) ≈ 3.85 years. For more realistic conditions—taking the actual quantum efficiency of 0.65 into account [7]—this number increases to 9.14 years. This also assumes the raster scanning is limited only by the acquisition time at each position, and that the time to translate between points is negligible.

## 4. Effects of Charge Transfer Inefficiency

During the readout process of the EMCCD camera, charge is transferred vertically through columns of the array to the readout register. This process does not occur with 100 % efficiency, i.e., the probability to transfer all the electrons from one pixel to the next less than unity. This concept is quantified by the Charge Transfer Efficiency (CTE), which may be very close to one (typically in excess of 0.9999996 [8]). Defect states in individual pixels are responsible for trapping charge, resulting in signal loss at the pixel of interest and vertical smearing. This effect has been studied extensively, particularly in the astronomy community, where ionization and bulk damage is caused by high-energy photons and particles in spacecraft [8-11].

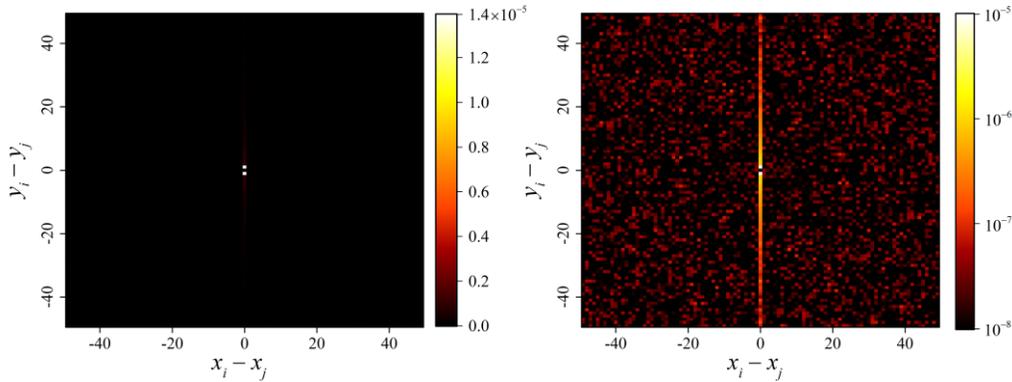

**Fig. S2 | Projection of $\Gamma_{ij}$ onto difference coordinates $[x_i - x_j, y_i - y_j]$.** Data from $3 \times 10^6$ thresholded frames of ($99 \times 99$) pixels measured with the camera shutter closed on a (a) linear and (b) logarithmic color scale. Pixel at [0, 0] is set to zero. Vertical line at $x_i - x_j = 0$ is due to imperfect vertical charge transfer during readout, resulting in artificial correlation between pixels in the same column.

This cross talk effect results in a correlation between one pixel and those in the same column, particularly in those immediately above and below. An example is shown Fig. S2, where $3 \times 10^6$ thresholded frames were measured with the camera shutter closed, such that all registered "clicks" originated from electronic noise. From these frames, $\Gamma_{ij}$ was calculated via Eq. (4) and projected onto difference coordinates. Nominally, since no biphotons reach the camera, the result should be zero. However, we clearly see a correlation, particularly between one pixel and the pixels directly above and beneath.